\documentclass[12pt]{iopart}

\newtheorem{thm}{Theorem}

 \newtheorem{prop}{Proposition}

\begin{document}

\title[Reciprocal link for a coupled Camassa-Holm type equation]{Reciprocal link for a coupled Camassa-Holm type equation}

\author{Nianhua Li, Jinshun Zhang, and Lihua Wu}

\address{School of Mathematics,
Huaqiao University,
Quanzhou, Fujian 362021, People's Republic of China}
\ead{linianh@hqu.edu.cn}
\begin{abstract}
A coupled Camassa-Holm type equation is linked to the first negative flow of a modified Drinfeld-Sokolov III hierarchy by a transformation of reciprocal type. Meanwhile the Lax pair and bi-Hamiltonian structure behaviors of this coupled Camassa-Holm type equation under change of variables are analyzed.
\end{abstract}

\section{Introduction}
The Camassa-Holm (CH) equation
\begin{equation}\label{ch}
m_t+um_x+2u_xm=0,\quad m=u-u_{xx}
\end{equation}
was proposed as a model for long waves in shallow water by the asymptotic approximation of Hamiltonian for Green-Naghdi equations in 1993 \cite{Holm}.
It is completely integrable with a Lax pair and associated bi-Hamiltonian structure \cite{Holm,hyman}, and is shown to be solvable by inverse scattering
transformation \cite{beal1,consta1,consta2}. Meanwhile the CH equation is linked to the first negative flow of the KdV hierarchy by a transformation
of reciprocal type \cite{fuch,wang1,lenel}. Furthermore, different from KdV equation, the CH equation admits peakon solutions \cite{Holm,hyman,Beals},
and we call the integrable equation possessing peakon solutions CH type equation.

Degasperis and Procesi \cite{Degasperis}, applying the method of asymptotic integrability, discover a new CH type equation
\begin{equation}
m_t+um_x+3u_xm=0,\quad m=u-u_{xx},
\end{equation}which is also integrable admitting a Lax representation as well as a bi-Hamiltonian structure, and is reciprocal linked to a negative flow of the Kaup-Kupershmidt hierarchy \cite{Deg2}. Besides the multi-peakon solutions of it are studied by inverse scattering approach \cite{Lun,lu2}.

Applying the tri-Hamiltonian duality approach \cite{fuch,olver}, two new CH type systems are proposed and the corresponding Lax pairs are given by Schiff \cite{schiff}. The first one is the modified CH equation (MCH) \cite{fuch,olver}
\begin{equation}
m_t+[(u^2-u_x^2)m]_x=0, \quad m=u-u_{xx},
\end{equation}which is later rediscovered by Qiao from two dimensional Euler equation \cite{qiao}. Furthermore, the MCH equation is related to the (modified) KdV hierarchy via a reciprocal transformation \cite{Hone}. The peakon and multi-component generalization of it are also researched (see e.g. \cite{Gui,xia}).
The second one is a two-component CH equation \cite{olver}
\begin{eqnarray}
&& m_t+um_x+2u_xm-\rho\rho_x = 0,\quad m=u-u_{xx},  \\
&& \rho_t+(\rho u)_x = 0,
\end{eqnarray}which is rediscovered from the Green-Naghdi equations and the peakon solutions of it in the short waves limit are constructed by Constantin \cite{constantin}.  Moreover it is reciprocal linked to the AKNS hierarchy \cite{Ming}.

Subsequently, the Novikov's equation \cite{Novikov}
\begin{equation}\label{novikov}
m_{t}+u^{2}m_{x}+3uu_{x}m=0,\quad m=u-u_{xx}.
\end{equation}is obtained in the symmetry classification of CH type equation, a Lax pair and bi-Hamiltonian structure are given by Hone and Wang \cite{Hone}, and the explicit formulas for the multi-peakon solutions of the Novikov's equation are calculated \cite{honeJ}. Especially the Novikov's equation is reciprocal connected to the first negative flow of Sawaka-Kotera hierarchy.  The Geng-Xue equation \cite{Geng}
 \begin{eqnarray}\label{novikov2}
&&m_{t}+3u_{x}vm+uvm_{x} =0, \nonumber\\
&&n_{t}+3v_{x}un+uvn_{x} =0,  \\
&&m=u-u_{xx},\quad n=v-v_{xx}. \nonumber
\end{eqnarray}is proposed as a generalization of the Novikov's equation admitting a Lax pair and a Hamiltonian structure later. Furthermore, the Geng-Xue equation is reciprocal connected to the first negative flow of the modified Boussinesq hierarchy \cite{Li}, and the peakon solutions of it are discussed \cite{ldk}.

Recently, Geng and Wang \cite{geng1} propose a new coupled CH type equation with cubic nonlinearity
\begin{eqnarray}\label{cubicCH}
&& v_t=2v_x(qr_x-q_xr)+2v(3qr_{xx}-q_{xx}r-q_xr_x-qr), \nonumber\\
&& w_t=2w_x(qr_x-q_xr)-2w(3q_{xx}r-qr_{xx}-q_xr_x-qr),\\
&& v=r_{xxx}-r_x,\quad w=q_{xxx}-q_x,\nonumber
\end{eqnarray}associated with a $4\times4$ matrix spectral problem
\begin{equation}\label{lax}
 \varphi_x=U\varphi,\quad \varphi=\left(
                           \begin{array}{c}
                             \varphi_1 \\
                             \varphi_2  \\
                             \varphi_3  \\
                             \varphi_4  \\
                           \end{array}
                         \right),\quad U=\left(
                                           \begin{array}{cccc}
                                             0 & 0 & 1 & 0 \\
                                             0 & 0 & 0 & 1 \\
                                             \frac{1}{4} & \lambda v & 0 & 0 \\
                                             \lambda w & \frac{1}{4} & 0 & 0 \\
                                           \end{array}
                                         \right).
\end{equation}
Bi-Hamiltonian structure and infinite sequences of conserved quantities as well as N-peakon solutions
for the new system are also considered by them.

The outline of this paper is as follows. In section 2, we construct a reciprocal transformation for the CH type system (\ref{cubicCH}), it shows that the transformed system is a constraint of the first negative flow of a modified Drinfeld-Sokolov (DS) III hierarchy. In section 3, the Hamiltonian structures of the coupled CH type hierarchy under the reciprocal transformation are given. In appendix, a reciprocal transformation for another system related to the problem (\ref{lax}) is also studied.
\section{Reciprocal transformation}
Since infinite sequences of conserved quantities for the integrable hierarchy of the spectral problem (\ref{lax}) are obtained \cite{geng1},  we may construct reciprocal transformations for the coupled CH type system (\ref{cubicCH}) which possesses the Lax representation
\begin{equation}\label{lax1}
 \varphi_{x}=U\varphi, \quad \varphi_{t}=V\varphi,
\end{equation}
where
\begin{equation*}
V=\left(
      \begin{array}{cccc}
       2q_{xx}r-k_1 & \frac{r_x}{\lambda}&4k_2 & -\frac{2r}{\lambda} \\
        -\frac{q_x}{\lambda} & k_1-2qr_{xx}  & \frac{2q}{\lambda} & 4k_2 \\
        q_{xx}r_x-q_xr_{xx}-k_2 &  4\lambda vk_2+\frac{2r_{xx}-r}{2\lambda} & 2qr_{xx}-k_1 & -\frac{r_x}{\lambda} \\
        4\lambda wk_2+\frac{q-2q_{xx}}{2\lambda} & q_{xx}r_x-q_xr_{xx}-k_2 & \frac{q_x}{\lambda} & k_1-2q_{xx}r\\
      \end{array}
    \right),
\end{equation*}herein $k_1=\frac{1}{2\lambda^2}+q_xr_x+qr, k_2=\frac{1}{2}(qr_x-q_xr)$.

Especially, notice that the equation (\ref{cubicCH}) implies a conservation law
\begin{equation*}
((wv)^{\frac{1}{4}})_t=(2(wv)^{\frac{1}{4}}(qr_x-q_xr))_x,
\end{equation*}which defines a reciprocal transformation via the relation
\begin{equation}\label{reci1}
dy=udx+2u(qr_x-q_xr)dt,\quad d\tau=dt,
\end{equation}where $u=(wv)^{\frac{1}{4}}$.

 Setting $h =(\frac{v}{w})^{\frac{1}{4}}$, then after a gauge
transformation $\varphi_1=hu^{-\frac{1}{2}}\phi$, the scalar form of spectral problem (\ref{lax}) can be transformed to
\begin{equation}\label{lax3}
(\partial_y^4+\partial_y m\partial_y+n)\phi=\lambda^2\phi,
\end{equation}where
\begin{eqnarray*}
 && m=-\frac{1}{2u^2}+\frac{u_y^2}{2u^2}-\frac{u_{yy}}{u}-6\frac{h_y^2}{h^2}+4\frac{h_{yy}}{h}, \\
 && n=-\frac{h_y}{h}m_{y}+\frac{h_y^2}{h^2}m+\frac{1}{2}m_{yy}
  -\frac{h_{yyyy}}{h}+36\frac{h_y^4}{h^4}-48\frac{h_y^2h_{yy}}{h^3}+6\frac{h_{yy}^2}{h^2}+10\frac{h_yh_{yyy}}{h^2}\\
 &&\hspace{0.8cm}+\frac{u_{yy}}{4u^3}+\frac{u_{yy}^2}{4u^2}+\frac{1}{16u^4}-\frac{u_y^2u_{yy}}{4u^3}+\frac{u_y^4}{16u^4}-\frac{u_y^2}{8u^4}.
\end{eqnarray*}

The relation between $m,n$ and $u,h$ may be related to factorization of Lax operator \cite{Fordy}. To begin with the factorization in \cite{guha},
\[
L=\partial_y^4+\partial_y m\partial_y+n=(\partial_y^2+\partial_y i -j)(\partial_y^2-i\partial_y-j),
\]where $m$ and $n$ satisfy a pair of generalized Miura system
\begin{equation*}
m=-(i_y+i^2+2j), \quad n=j^2-j_{yy}-(ij)_y.
\end{equation*}
In fact $L$ may be further decomposed as
\begin{equation*}
L=(\partial_y-b_1+a_1)(\partial_y-b_1-a_1)(\partial_y+b_1+a_1)(\partial_y+b_1-a_1),
\end{equation*}
with $i,j$ satisfying
\begin{eqnarray*}
  i=-2b_1,\quad j=a_{1y}+a_1^{2}-b_{1y}-b_1^{2},
\end{eqnarray*}
where $a_1=\frac{u_{y}+1}{2u},\ b_1=\frac{h_y}{h}$.

It is not difficult to find that the spectral problem (\ref{lax3}) is which of the DS III system \cite{metin,sokolov}
and the  system (\ref{cubicCH}) in the new variable is nothing but a constraint of the first negative flow of the DS III hierarchy.
However, detailed calculation shows that we should rewrite the spectral problem (\ref{lax3}) as matrix form in order to obtain the transformed system of
(\ref{cubicCH}) and its Lax pair.

 To this end, on the one hand, eliminating $\varphi_3,\varphi_4$ and denoting $s=\frac{u}{h^2}$,
the spectral problem (\ref{lax}) is transformed to
\begin{eqnarray*}
&&\phi_{yy}+(\frac{u_y}{u}-\frac{s_y}{s})\phi_y+(\frac{3s_y^2}{4s^2}-\frac{s_{yy}}{2s}-\frac{u_ys_y}{2us}-\frac{1}{4u^2})\phi=\lambda\psi,\\
&&\psi_{yy}+(\frac{s_y}{s}-\frac{u_y}{u})\psi_y+(\frac{s_{yy}}{2s}-\frac{s_y^2}{4s^2}-\frac{u_ys_y}{2us}-\frac{u_{yy}}{u}+\frac{u_y^2}{u^2}-\frac{1}{4u^2})\psi=\lambda\phi,
\end{eqnarray*}where $\psi=us^{-\frac{1}{2}}\varphi_2$ (here and in the sequel, we use $u,s$ instead of $u,h$ for convenience).
Then the two new potentials $i,j$ are obtained
\begin{equation*}
i=\frac{s_y}{s}-\frac{u_y}{u}, \quad j=\frac{1}{4u^2}+\frac{u_ys_y}{2us}-\frac{3s_y^2}{4s^2}+\frac{s_{yy}}{2s}.
\end{equation*} Setting
$\Phi=(\psi,
    \psi_y,
    \phi,
    \phi_y)^{T}$, then the Lax pair (\ref{lax1}) of the system (\ref{cubicCH}) is transformed to
\begin{equation}\label{tran1}
\Phi_y= \left(
             \begin{array}{cccc}
               0 & 0 & 1 & 0 \\
               0 & 0 & 0 & 1 \\
               j & \lambda & i & 0 \\
               \lambda & j-i_y & 0 & -i \\
             \end{array}
           \right)\Phi,\quad
           \Phi_\tau=\left(
               \begin{array}{cccc}
                 -\frac{1}{2\lambda^2} & \frac{g_y-2gi}{\lambda} & 0 & -\frac{2g}{\lambda} \\
                 -\frac{f_y+2fi}{\lambda} & \frac{1}{2\lambda^2}& \frac{2f}{\lambda} & 0 \\
                 -2g & \frac{A_2}{\lambda}& -\frac{1}{2\lambda^2} & -\frac{g_y}{\lambda} \\
                 \frac{A_3}{\lambda} & 2f & \frac{f_y}{\lambda} & \frac{1}{2\lambda^2} \\
               \end{array}
             \right)\Phi,
\end{equation}where $f=q\frac{u^2}{s},g=rs$ with
\begin{eqnarray*}
  A_2 &=& g_{yy}-2g_yi-2gj, \\
  A_3 &=& -f_{yy}-2(fi)_y+2fj.
\end{eqnarray*}

On the other hand, under the transformation (\ref{reci1}), the system (\ref{cubicCH}) is transformed to
\begin{eqnarray*}
  i_\tau &=& -2(rs+q\frac{u^2}{s}),\quad \quad \quad \quad \quad \quad \ u^2s^{-1}=\partial_y u\partial_y u\partial_y r-\partial_y r, \\
  j_\tau &=& -3(rs)_y+2rsi+\frac{u^2}{s}(q_y+q\frac{s_y}{s}), \quad s=\partial_y u\partial_y u\partial_y q-\partial_y q.
\end{eqnarray*}
Furthermore, above system may be reformed as
\begin{eqnarray}\label{md1}
  i_\tau &=& -2(f+g),\quad \quad \quad \quad \quad \  F_1=-1, \\\label{md2}
  j_\tau &=& f_y-3g_y+2i(f+g),\quad F_2=-1,
\end{eqnarray}
where
\begin{eqnarray*}
&&F_1=3ig_{yy}-g_{yyy}+(i_y-2i^2+4j)g_y+(2j_y-4ij)g, \\
&&F_2=(4ij+2j_y-2i_{yy}-4ii_y)f+(4j-5i_y-2i^2)f_y-3if_{yy}-f_{yyy}.
\end{eqnarray*}
Then it is easy to verify that the compatibility of the transformed Lax pair (\ref{tran1}) is just the transformed system (\ref{md1}-\ref{md2}),
 so the system (\ref{cubicCH}) and its Lax pair (\ref{lax1}) are transformed to (\ref{md1}-\ref{md2}) and (\ref{tran1}) respectively.

Now, we will show that the transformed system (\ref{md1}-\ref{md2}) is a constraint of the first negative flow of the modified DS III hierarchy.
The first negative flow of the modified DS III hierarchy may be formed as (see Appendix)
\begin{eqnarray}\label{neg1}
  i_\tau &=& -2(f+g), \quad \quad \quad \quad \quad \ G_1=0,\\\label{neg2}
  j_\tau &=& f_y-3g_y+2i(f+g),\quad G_2=0,
\end{eqnarray}
where
\begin{eqnarray*}
  G_1 &=& \frac{1}{4}(2(F_1+F_2)_y+\partial_y i\partial_y^{-1}(F_1-F_2)),\\
  G_2 &=& \frac{1}{4}((\partial_y^2-i\partial_y)(F_1+F_2)+(-\frac{1}{2}\partial_y^3+\frac{i}{2}\partial_y^2+j\partial_y+\partial_y j)\partial_y^{-1}(F_1-F_2)).
\end{eqnarray*}
To see the connection between (\ref{md1}-\ref{md2}) and (\ref{neg1}-\ref{neg2}), it is clear that the following identity holds:
\begin{equation*}
\left(
  \begin{array}{c}
    G_1 \\
    G_2 \\
  \end{array}
\right)=\frac{1}{2}\left(
                     \begin{array}{cc}
                       \partial_y+\frac{i}{2}+\frac{i_y}{2}\partial_y^{-1} &  \partial_y-\frac{i}{2}-\frac{i_y}{2}\partial_y^{-1} \\
                        \frac{1}{4}\partial_y^2-\frac{i}{4}\partial_y+j+\frac{j_y}{2}\partial_y^{-1} & \frac{3}{4}\partial_y^2-\frac{3i}{4}\partial_y-j-\frac{j_y}{2}\partial_y^{-1} \\
                     \end{array}
                   \right)\left(
                  \begin{array}{c}
                    F_1 \\
                    F_2\\
                  \end{array}
                \right),
\end{equation*}(here all integration constants are assumed to be zero).
This leads to the fact that the system (\ref{md1}-\ref{md2}) may be regarded as a reduction of the first negative flow of the modified DS III hierarchy (\ref{neg1}-\ref{neg2}).

\section{The Hamiltonian structure behavior under the transformation}
Let us define ${\cal E}=\partial^3-\partial,\ \theta=(v,w)^{T}$ and $v^{(n)}=\frac{\partial^{n}v}{\partial x^n}$, then the coupled system (\ref{cubicCH}) can be written as a bi-Hamiltonian structure
\begin{equation}
\left(
  \begin{array}{c}
    v \\
    w \\
  \end{array}
\right)_t={\cal J}_2\left(
                      \begin{array}{c}
                        \frac{\delta}{\delta v} \\
                        \frac{\delta}{\delta w} \\
                      \end{array}
                    \right)H_0={\cal J}_1\left(
                      \begin{array}{c}
                        \frac{\delta}{\delta v} \\
                        \frac{\delta}{\delta w} \\
                      \end{array}
                    \right)H_1
\end{equation}where
\begin{eqnarray*}
&&{\cal J}_2=-{\cal E}\sigma_1,\\
&&{\cal J}_1=-2\sigma_3\theta\partial^{-1}(\sigma_3\theta)^{T}-2(\theta\partial+\partial \theta){\cal E}^{-1}(\theta\partial+\partial \theta)^{T},
\end{eqnarray*}and
\begin{eqnarray*}
  H_0 &=& \frac{1}{2}\int (v[2q^2r_{xx}-2qq_xr_x+q_x^2r-q^2r]+w[2rq_xr_x-2q_{xx}r^2-r_x^2q+r^2q]) dx, \\
  H_1 &=& \frac{1}{2}\int (wr-qv) dx,
\end{eqnarray*} herein $\sigma_1=\left(
                             \begin{array}{cc}
                               0 & 1 \\
                               1 & 0 \\
                             \end{array}
                           \right)$ and $\sigma_3=\left(
                                              \begin{array}{cc}
                                                1 & 0 \\
                                                0 & -1 \\
                                              \end{array}
                                            \right)
$ are the standard Pauli matrices.

The corresponding structures such as the recursion operator, bi-Hamiltonian structure, conserved quantities between the two connected equations can be generated by the reciprocal transformation (\ref{reci1}), and in the following, the bi-Hamiltonian structure for the coupled CH type hierarchy under the change of variables $(x,v,w)\rightarrow (y,i,j)$
\begin{eqnarray*}
\left\{\begin{array}{rl}
&y=P(x,v^{(n)},w^{(n)})=\int^x_{-\infty} (wv)^{\frac{1}{4}}dx=\partial^{-1} (wv)^{\frac{1}{4}}, \\
&i(y)=Q_1(x,v^{(n)},w^{(n)})=-\frac{1}{2}(wv)^{-\frac{5}{4}}(wv_x-w_xv),\\
&j(y)=Q_2(x,v^{(n)},w^{(n)})\\
&\hspace{0.8cm}=\frac{1}{64}(wv)^{-\frac{5}{4}}(16w^2v^2-33v^2w_x^2+6ww_xvv_x+7w^2v_x^2+24v^2ww_{xx}-8w^2vv_{xx})\end{array} \right.
\end{eqnarray*}may be given.

Let $\vartheta=(i,j)^{T}$, following the ideal in \cite{Bru}, an implicit function $B(\theta,\vartheta)=0$ may be defined, then
\begin{equation*}
B_{\theta}\theta_t+B_{\vartheta}\vartheta_t= 0,
\end{equation*} where $B_{\theta},B_{\vartheta}$ are corresponding Frech\'et derivatives for the vector variables, so we get
\begin{equation*}
\vartheta_t=-T_1\theta_t,\quad T_1=B_{\vartheta}^{-1}B_{\theta}.
\end{equation*} It is easy to find that $B_{\vartheta}$ is the identity matrix, therefore we may obtain
\begin{equation}\label{T1}
T_1=\left(
                                        \begin{array}{cc}
                                         i_yP'[v]-Q_1'[v] &i_yP'[w]-Q_1'[w] \\
                                         j_yP'[v]-Q_2'[v] &j_yP'[w]-Q_2'[w] \\
                                        \end{array}
                                      \right).
\end{equation}
Then as we know that
\begin{eqnarray}\label{HH1}
&&\theta_t={\cal J}(v,w)\frac{\delta H}{\delta\theta}={\cal J}(v,w)E_{\theta}h,\\\label{HH2}
&&\vartheta_t=\tilde{{\cal J}}(i,j)\frac{\delta \tilde{H}}{\delta\vartheta}=\tilde{{\cal J}}(i,j)E_{\vartheta}\tilde{h},
\end{eqnarray}where
\begin{equation*}
H=\int h(x,v^{(n)},w^{(n)})dx,\quad \tilde{H}=\int \tilde{h}(y,i^{(n)},j^{(n)})dy,
\end{equation*}
and $E_{\theta}, E_{\vartheta}$ are the Euler operators. In order to connect the Hamiltonian structures of the two evolution equations, we need
the action on Euler operator under a change of variables which is related by (see \cite{olver2}, Exercise 5.49)
\begin{equation*}
E_{\theta}h=T_2E_{\vartheta}\tilde{h},
\end{equation*}
herein
\begin{equation}\label{T2}
T_2=\left(
\begin{array}{cc}
Q'^{*}_{1,v}(D_x P)-P'^{*}_{v}(D_x Q_1) & Q'^{*}_{2,v}(D_x P)-P'^{*}_{v}(D_x Q_2) \\
Q'^{*}_{1,w}(D_x P)-P'^{*}_{w}(D_x Q_1) & Q'^{*}_{2,w}(D_x P)-P'^{*}_{w}(D_x Q_2) \\
\end{array}
\right).
\end{equation}
We are now in a position to state our main results:
\begin{thm}\cite{Bru} The Hamiltonian structures of two evolution equations (\ref{HH1}) and (\ref{HH2}) which are related by $B(\theta,\vartheta)$  are linked as
\begin{equation}
  \tilde{{\cal J}}_{k}=-T_1{\cal J}_{k}T_2\ (k=1,2), \quad \left(
                         \begin{array}{c}
                           \frac{\delta }{\delta v} \\
                           \frac{\delta}{\delta w} \\
                         \end{array}
                       \right)H(v,w)=T_2\left(
                         \begin{array}{c}
                           \frac{\delta}{\delta i} \\
                           \frac{\delta}{\delta j} \\
                         \end{array}
                       \right)\tilde{H}(i,j),
\end{equation}where $T_1$ and  $T_2$ are given by (\ref{T1}) and (\ref{T2}) accordingly.
\end{thm}

In the following, we will indicate the explicit formula for the transformed Hamiltonian operators.
It is easy to find that
\begin{eqnarray*}
&&P'_{v}=\frac{1}{4}\partial^{-1}(wv)^{-\frac{3}{4}}w=\frac{1}{4}\partial_y^{-1}\frac{1}{v},\quad \quad \quad \quad \ P'_{w}=\frac{1}{4}\partial_y^{-1}\frac{1}{w},\\
&&P'^{*}_{v}=-\frac{1}{4}(wv)^{-\frac{3}{4}}w\partial^{-1}=-\frac{u}{4v}\partial_y^{-1}\frac{1}{u},\quad \quad P'^{*}_{w}=-\frac{u}{4w}\partial_y^{-1}\frac{1}{u}.
\end{eqnarray*}
Then through tedious calculations and change of variables, we obtain
\begin{eqnarray*}
&&Q'_{1,v}=-\partial_y\frac{1}{2v}-\frac{i}{4v},\quad \quad \quad \quad \quad \quad \quad  \  Q'_{1,w}=\partial_y\frac{1}{2w}-\frac{i}{4w},\\
&&Q'_{2,v}=(-\frac{j}{2}-\frac{1}{8}\partial_y^2+\frac{i}{8}\partial_y)\frac{1}{v},\quad \quad \quad \quad Q'_{2,w}=(-\frac{j}{2}+\frac{3}{8}\partial_y^2-\frac{3i}{8}\partial_y)\frac{1}{w},\\
\end{eqnarray*}and
\begin{eqnarray*}
&&Q'^{*}_{1,v}(D_x P)=\frac{u}{4v}(2\partial_y-i),\quad \quad \quad \quad \quad  \quad \   Q'^{*}_{1,w}(D_x P)=-\frac{u}{4w}(2\partial_y+i),\\
&&Q'^{*}_{2,v}(D_x P)=-\frac{u}{v}(\frac{j}{2}+\frac{1}{8}\partial_y^2+\frac{1}{8}\partial_y i),\quad \quad Q'^{*}_{2,w}(D_x P)=\frac{u}{w}(-\frac{j}{2}+\frac{3}{8}\partial_y^2+\frac{3}{8}\partial_y i).\\
\end{eqnarray*}
A direct computation shows
\begin{eqnarray*}
  T_1 &=& \frac{1}{4}\left(
            \begin{array}{cc}
              (\partial_y i \partial_y^{-1}+2\partial_{y})\frac{1}{v}& (\partial_y i \partial_y^{-1}-2\partial_{y})\frac{1}{w} \\
               (j_y\partial_{y}^{-1}+2j+\frac{1}{2}\partial_y^2-\frac{i}{2}\partial_y)\frac{1}{v}& (j_y\partial_{y}^{-1}+2j-\frac{3}{2}\partial_y^2+\frac{3i}{2}\partial_y)\frac{1}{w} \\
            \end{array}
          \right), \\ \\
T_2 &=&  \frac{1}{4}\left(
            \begin{array}{cc}
               \frac{u}{v}(2\partial_y-i+\partial_y^{-1}i_y)& \frac{u}{v}(-\frac{1}{2}\partial_y^2-\frac{1}{2}\partial_y i-2j+\partial_y^{-1}j_y) \\
               \frac{u}{w}(-2\partial_y-i+\partial_y^{-1}i_y) & \frac{u}{w}(\frac{3}{2}\partial_y^2+\frac{3}{2}\partial_y i-2j+\partial_y^{-1}j_y) \\
            \end{array}
          \right).
\end{eqnarray*}
We are now in a position to obtain the Hamiltonian operators $\tilde{{\cal J}}_1,\tilde{{\cal J}}_2$.
\begin{prop}Under change of variables,
\begin{equation*}
  \frac{1}{v}{\cal E}\frac{u}{w}=\Theta_1=\partial_y^3-3i\partial_y^2+(2i^2-i_y-4j)\partial_y+4ij-2j_y.
\end{equation*}
\end{prop}
It follows from the proposition 1 that
\begin{equation*}
\frac{1}{w}{\cal E}\frac{u}{v}=-\Theta_1^{*}=\partial_y^3+3i\partial_y^2+(2i^2+5i_y-4j)\partial_y+2i_{yy}-2j_y+4ii_y-4ij,
\end{equation*}
Hence  we deduce that
\begin{equation}
\tilde{{\cal J}}_1=-\frac{1}{16}\Lambda\left(
                                      \begin{array}{cc}
                                        0 & \Theta_1 \\
                                        -\Theta_1^{*} & 0 \\
                                      \end{array}
                                    \right)\Lambda^{*},
\end{equation}where
\begin{equation*}
\Lambda=\left(
                                \begin{array}{cc}
                                  \partial_y i\partial_y^{-1}+2\partial_y & \partial_y i\partial_y^{-1}-2\partial_y \\
                                  j_y\partial_{y}^{-1}+2j+\frac{1}{2}\partial_y^2-\frac{i}{2}\partial_y& j_y\partial_{y}^{-1}+2j-\frac{3}{2}\partial_y^2+\frac{3i}{2}\partial_y \\
                                \end{array}
                              \right).
\end{equation*}

\begin{prop}Under change of variables,
\begin{equation*}
  \frac{1}{u^2}{\cal E}\frac{1}{u}=\Theta_2=\partial_y^3+(i_y-\frac{1}{2}i^2-2j)\partial_y+\partial_y(i_y-\frac{1}{2}i^2-2j).
\end{equation*}
\end{prop}
Using the proposition 2, a direct calculation shows
\begin{eqnarray*}
&&\tilde{{\cal J}}_2=\frac{1}{2}\left(
                                \begin{array}{c}
                                  2 \\
                                  \partial_y-i \\
                                \end{array}
                              \right)\partial_y\left(
                                \begin{array}{c}
                                  2 \\
                                  \partial_y-i \\
                                \end{array}
                              \right)^{*}-\frac{1}{2}\Theta_2\left(
                                                               \begin{array}{cc}
                                                                 0 & 0 \\
                                                                 0 & 1 \\
                                                               \end{array}
                                                             \right)\\
&&\hspace{0.6cm}=\left(
     \begin{array}{cc}
       2\partial_y & -(\partial_y^2+\partial_y i) \\
       \partial_y^2-i\partial_y & j\partial+\partial j-(\partial_y-i)\partial_y(\partial_y+i) \\
     \end{array}
   \right)                            .
\end{eqnarray*}
Since the compatible Hamiltonian operators are gotten, a recursion operator $\tilde{{\cal J}}_2\tilde{{\cal J}}_1^{-1}$ is obtained which is also the recursion operator of the modified DS III hierarchy through complicated calculation (compare with the recursion operator of modified DS III hierarchy in appendix), besides the $n$th equation of the coupled CH type hierarchy may be mapped similarly.

\noindent
{\bf ACKNOWLEDGMENTS}

This work is partially supported by the National Natural Science Foundation of China (Grant Nos. 11401572 and 11401230) and the Initial Founding of Scientific Research for the introduction of talents of Huaqiao University (Project No. 14BS314).

\appendix
\renewcommand\theequation{\Alph{section}\arabic{equation}}

\section{THE FIRST NEGATIVE FLOW OF THE MODIFIED DS III HIERARCHY}

Firstly, we will derive the recursion operator of the modified DS III hierarchy.
Actually, the bi-Hamiltonian operators of the DS III hierarchy \cite{metin,mihaklov} are
\begin{eqnarray*}
&&P^1=\left(
      \begin{array}{cc}
        0 & 4\partial_y \\
        4\partial_y & 3\partial_y^3+\partial_y m+m\partial_y \\
      \end{array}
    \right),\\
&&P^2=\left(
      \begin{array}{cc}
        5\partial_y^3+m\partial_y+\partial_y m & \frac{3}{2}\partial_y^5+\frac{3}{2}\partial_y^2m\partial_y+4n\partial_y+3n_y \\
        \frac{3}{2}\partial_y^5+\frac{3}{2}\partial_y m\partial_y^2+4n\partial_y+n_y & P^{2}_{22} \\
      \end{array}
    \right),
\end{eqnarray*}
where $P^{2}_{22}=\frac{1}{2}(\partial_y^7+\partial_y(\partial_y^3m+m\partial_y^3+m\partial_y m+n\partial_y+\partial_y n)\partial_y)+\partial_y^3n+n\partial_y^3+\partial_y mn+mn\partial_y$,
then the recursion operator of the DS III hierarchy is $R=P^2(P^1)^{-1}$. Notice that
\begin{equation*}
\left(
                                                             \begin{array}{c}
                                                               m \\
                                                               n \\
                                                             \end{array}
                                                           \right)=\Omega(i,j)=\left(
                                                                                 \begin{array}{c}
                                                                                    -i_y-i^2-2j\\
                                                                                    j^2-j_{yy}-(ij)_y \\
                                                                                 \end{array}
                                                                               \right),
\end{equation*}
therefore the recursion operator of the modified DS III hierarchy is $\tilde{R}=\Omega'^{-1}R\Omega'$.

Secondly, let us introduce the first negative flow of the modified DS III hierarchy from the formal Lax pair
\begin{equation*}
\Phi_y=\left(
                                                 \begin{array}{cc}
                                                   0 & I\\
                                                   A & B \\
                                                 \end{array}
                                               \right)\Phi, \quad \Phi_\tau=\left(
                                                 \begin{array}{cc}
                                                   K^{1} & K^{2}\\
                                                   K^{3} & K^{4} \\
                                                 \end{array}
                                               \right)\Phi,
\end{equation*}
where
\begin{eqnarray*}
 A=\left(
                  \begin{array}{cc}
                    j & \lambda \\
                   \lambda & j-i_y \\
                  \end{array}
                \right),\quad B=\left(
                                  \begin{array}{cc}
                                    i & 0 \\
                                    0 & -i \\
                                  \end{array}
                                \right),
\end{eqnarray*}$I$ is the identity matrix, and $K^{k}=(K^{k})_{2\times 2},(k=1,...,4)$. The zero-curvature equation yields
\begin{eqnarray*}
&&K^{1}=K^{4}-K^{2}_y-K^{2}B, \\
&&K^{3}=K^{4}_y-K^{2}_{yy}-(K^{2}B)_y+K^{2}A,\\
&&A_\tau=K^{3}_y-AK^{1}-BK^{3}+K^{4}A,\\
&&B_\tau=K^{4}_y-AK^{2}-BK^{4}+K^{3}+K^{4}B,
\end{eqnarray*}
which imply that
\begin{eqnarray*}
&&(i-\partial_y)(2K^4_{12}-K^2_{12y})+\lambda(K^2_{22}-K^2_{11})=0, \\
&&(i+\partial_y)(2K^4_{21}-K^2_{21y})+\lambda(K^2_{22}-K^2_{11})=0, \\
&&K^4_{11}+K^4_{22}-\frac{1}{2}[(\partial_y+i)K^2_{11}+(\partial_y-i)K^2_{22}]=0,\\
&&S_1-S_2-2\lambda(K^2_{11}+K^2_{22})_y=0, \\
&&S_1+S_2+\lambda[2K^4_{22}-2K^4_{11}+K^2_{11y}-K^2_{22y}+2i(K^2_{11}+K^2_{22})]=0, \\
&&[\frac{1}{2}(\partial_y^3+\partial_y^2 i-i\partial_y^2-i\partial_y i)-j\partial_y-\partial_y j](K^2_{11}-K^2_{22})
  +\lambda(K^2_{12y}-K^2_{21y}+2K^4_{21}-2K^4_{12})=0,
\end{eqnarray*}
and
\begin{eqnarray}\label{fir1}
&&i_\tau=\lambda(K^2_{12}-K^2_{21})+\lambda^{-1}M_1, \\\label{fir2}
&&j_\tau=\lambda[\frac{1}{2}(3K^2_{12}+K^2_{21})_y+i(K^2_{21}-K^2_{12})]+\lambda^{-1}M_2,
\end{eqnarray}
where
\begin{eqnarray*}
  S_1 &=&(\frac{1}{2}\partial_y^3-i\partial_y^2-j\partial_y-\partial_y j-\frac{1}{2}\partial_y i\partial_y+i^2\partial_y+2ij)K^2_{12}, \\
  S_2 &=&-(\frac{1}{2}\partial_y^3+\partial_y^2 i-j\partial_y-\partial_y j+\frac{1}{2}\partial_y i\partial_y+2i\partial_y i-i^2\partial_y-2ij)K^2_{21}, \\
  M_1 &=& \frac{1}{4}[2(S_1+S_2)_y+\partial_y i\partial_y^{-1}(S_1-S_2)], \\
  M_2 &=& \frac{1}{4}[(\partial_y^2-i\partial_y)(S_1+S_2)+(-\frac{1}{2}\partial_y^3+\frac{i}{2}\partial_y^2+j\partial_y+\partial_y j)\partial_y^{-1}(S_1-S_2)].
\end{eqnarray*}
Then the first negative flow of the modified DS III hierarchy can be obtained by taking $K^2_{12}=-2g\lambda^{-1}, K^2_{21}=2f\lambda^{-1}$ in (\ref{fir1}-\ref{fir2})
\begin{equation}\label{flow}
\left(
  \begin{array}{c}
    i \\
    j \\
  \end{array}
\right)_\tau=
{\cal K}\left(
                             \begin{array}{c}
                               f \\
                               g \\
                             \end{array}
                           \right),\quad {\cal J}\left(
                                             \begin{array}{c}
                                               F_1 \\
                                               F_2 \\
                                             \end{array}
                                           \right)=\left(
                                                     \begin{array}{c}
                                                       0 \\
                                                       0 \\
                                                     \end{array}
                                                   \right),
\end{equation}
where
\begin{eqnarray*}
{\cal K} &=& \left(
             \begin{array}{cc}
               -2 & -2 \\
               \partial_y+2i & -3\partial_y+2i \\
             \end{array}
           \right),\\
{\cal J} &=&  \frac{1}{2}\left(
                     \begin{array}{cc}
                       \partial_y+\frac{i}{2}+\frac{i_y}{2}\partial_y^{-1} &  \partial_y-\frac{i}{2}-\frac{i_y}{2}\partial_y^{-1} \\
                        \frac{1}{4}\partial_y^2-\frac{i}{4}\partial_y+j+\frac{j_y}{2}\partial_y^{-1} & \frac{3}{4}\partial_y^2-\frac{3i}{4}\partial_y-j-\frac{j_y}{2}\partial_y^{-1} \\
                     \end{array}
                   \right).
\end{eqnarray*}
and
\begin{equation*}
\left(
  \begin{array}{c}
    F_1 \\
    F_2 \\
  \end{array}
\right)=\Theta\left(
                 \begin{array}{c}
                   f \\
                   g \\
                 \end{array}
               \right),\quad \quad \quad
\Theta=\left(
          \begin{array}{cc}
            0 & -\Theta_1 \\
             \Theta_1^{*} & 0 \\
          \end{array}
        \right).
\end{equation*}
Straightforward calculation shows that ${\cal J}\Theta{\cal K}^{-1}=-\tilde{R}$, so the system (\ref{flow}) is just the first negative flow of the modified DS III hierarchy.

\section{RECIPROCAL TRANSFORMATION OF A TWO-COMPONENT CH TYPE EQUATION}

In fact another two-component system
\begin{eqnarray}\label{coupleCH}
&&v_t=4vq_x+2v_xq+2vr,\nonumber\\
&&w_t=4wq_x+2w_xq-2wr,\\
&&v=q_{xx}-q+r_x,\quad w=q_{xx}-q-r_x,\nonumber
\end{eqnarray}is also given by Geng and Wang \cite{geng1}, this system and (\ref{cubicCH}) share the same spectral problem,  but auxiliary problem here is
\begin{eqnarray}
 \varphi_{t}=\left(
      \begin{array}{cccc}
        r-q_x & 0 & 2q & \frac{1}{\lambda} \\
        0 & -r-q_x & \frac{1}{\lambda} & 2q \\
        \frac{1}{2}(v+w+q)-q_{xx} &  2\lambda vq+\frac{1}{4\lambda} & r+q_x & 0 \\
        2\lambda wq+\frac{1}{4\lambda} & \frac{1}{2}(v+w+q)-q_{xx} & 0 & q_x-r \\
      \end{array}
    \right)\varphi.\nonumber
\end{eqnarray}
As points out in \cite{geng1}, the two-component CH system (\ref{coupleCH}) possesses a closed 1-form
\begin{equation*}
 \omega=(vw)^{\frac{1}{4}}dx+2q(vw)^{\frac{1}{4}}dt,
\end{equation*} which defines a reciprocal transformation as
\begin{equation}\label{trs2}
dy=udx+2qudt, \quad d\tau=dt.
\end{equation}
Proceeding as before we obtain, the CH system (\ref{coupleCH}) is reciprocally transformed to
\begin{eqnarray}\label{coutr1}
  i_{\tau} &=& s-\frac{u^2}{s}, \quad \quad \quad \quad \quad \quad i=\frac{s_y}{s}-\frac{u_y}{u},\\\label{coutr2}
  j_{\tau} &=& \frac{1}{2}s_y+\frac{u_y}{u}s+\frac{u^2s_y}{2s^2}, \quad  j=\frac{1}{4u^2}+\frac{u_ys_y}{2us}-\frac{3s_y^2}{4s^2}+\frac{s_{yy}}{2s},
\end{eqnarray}and the transformed Lax pair is
\begin{eqnarray*}
&&\Phi_y=\left(
             \begin{array}{cccc}
               0 & 0 & 1 & 0 \\
               0 & 0 & 0 & 1 \\
               j & \lambda & i & 0 \\
               \lambda & j-i_y & 0 & -i \\
             \end{array}
           \right)\Phi, \quad \Phi_\tau=\left(
               \begin{array}{cccc}
                 0 & \frac{s}{\lambda}(\frac{s_y}{2s}-\frac{u_y}{u})& 0 & \frac{s}{\lambda} \\
                 -\frac{u^2s_y}{2\lambda s^2} & 0 & \frac{u^2}{\lambda s} & 0 \\
                 s & \frac{A_1}{\lambda}& 0 & \frac{s_y}{2\lambda} \\
                 \frac{u^2A_1}{s^2\lambda} & \frac{u^2}{s} & (\frac{u^2}{2\lambda s})_y& 0 \\
               \end{array}
             \right)\Phi,
\end{eqnarray*}where $A_1=s(\frac{1}{4u^2}+\frac{s_y^2}{4s^2}-\frac{u_ys_y}{2su})$.

It is not difficult to find that the transformed system (\ref{coutr1}-\ref{coutr2}) is a constraint of the first negative flow of modified DS III hierarchy,
but the linear differential polynomials for $s,\frac{u^2}{s}$ like $f,g$ in $F_1,F_2$ are difficult to analyse, therefore the concrete relation between (\ref{coutr1}-\ref{coutr2})
and the first negative flow of the modified DS III hierarchy is open.

\end{document}